\documentclass{article}
\usepackage{amssymb}
\usepackage{amsfonts}
\usepackage{amsmath}
\usepackage{cite}
\usepackage{hyperref}
\usepackage{graphicx}

\providecommand{\keywords}[1]{\textbf{\text{Keywords:}} #1}

\makeatletter
\newcommand*\bigcdot{\mathpalette\bigcdot@{.5}}
\newcommand*\bigcdot@[2]{\mathbin{\vcenter{\hbox{\scalebox{#2}{$\m@th#1\bullet$}}}}}
\makeatother

\begin{document}

\title{Decays of $W$ bosons: possible mixing of spin $0$ and spin $1$ states}
\author{Andrzej Okni\'{n}ski \\
Chair of Mathematics and Physics, Politechnika \'{S}wi\c{e}tokrzyska,\\
Al. 1000-lecia PP 7, 25-314 Kielce, Poland}
\maketitle

\begin{abstract}
We study decays of a spin $1$ boson within the formalism of Hagen-Hurley
equations. Such particle can decay into two spin $\frac{1}{2}$ particles, a
Weyl neutrino and a massive fermion, whose spins can couple to $S=0$ or $S=1$%
. Since spin $0$ and spin $1$ bosons can be described by the Dirac equation
within the same representation of $\gamma ^{\mu }$ matrices mixing of $S=0$
and $S=1$ states is possible. We argue that the Hagen-Hurley equations
describe $W$ boson with spin $S\in 0\oplus 1$ space and analyse mixed beta
decays as well as top quark decays from this perspective. 
\smallskip

\noindent{\keywords{$W$ boson, top quark, beta decay}}
\end{abstract}

\section{Introduction}
\label{introduction}

Recently, we have split the $7\times 7$ Hagen-Hurley equations describing
spin $1$ particles in the interacting case, into two Dirac equations with
nonstandard solutions \cite{Okninski2016}. More exactly, we have obtained
the Weyl and the Dirac equations, describing two spin $\frac{1}{2}$
particles, a Weyl neutrino and a massive fermion, with spins coupling to $%
S=0 $ or $S=1$ \cite{Okninski2016}.

In our recent paper we have demonstrated that spin $\frac{1}{2}$ fermions as
well as spin $0$ Duffin-Kemmer-Petiau bosons as well as spin $1$
Hagen-Hurley bosons can be described by the Dirac equation within the same
representation of $\gamma ^{\mu }$ matrices \cite{Okninski2015}. We show
that within this formalism a transition between integer spins (mixing) of
these bosons is possible. This finding supports hypothesis that a particle
obeying the the $7\times 7$ Hagen-Hurley equations has spin in $0\oplus 1$
space.

We have suggested that these solutions describe decay of a virtual $W$ boson
in beta decay. Although the spin is conserved in the process of beta decay,
spin conservation of the $W$ boson as a virtual particle may be disputable.
It is argued that in pion decay as well as in the Fermi mechanism the $W$
boson is in $S_{z}=0$ state (is longitudinally polarized) and behaves as
spin $0$ particle. On the other hand, assuming that the virtual $W$ boson
obeys the Hagen-Hurley equations, we propose that spin of the $W$ boson
belongs to $0\oplus 1$ space. Therefore, in this work we are looking for a
mechanism permitting transition between spins $0$ and $1$ explaining
existence of Gamow-Teller and Fermi competing mechanisms of beta decay.

New information may be provided by the top quark decay. Since the mass of
the top quark is greater than mass of the $W$ boson the $W$ boson is a real
particle in $t\longrightarrow bW$ decay. Therefore, it can be expected that
the spin of the $W$ boson is $S=1$ and conserved. Alternatively, also the
real $W$ boson is described by the Hagen-Hurley equations and, due to spin $%
0 $ and $1$ mixing, has only partly determined spin, i.e. $S\in 0\oplus 1$
space.

In the next Section we describe shortly generalized solutions of the Dirac
equation using the same conventions and notation as in \cite{Okninski2015}.
Transformation, describing transition between integer spins and making spin
mixing possible, is also constructed. Then, in the next two Sections the
Gamow-Teller and Fermi mechanisms as well as top quark\ decay are analysed
in the light of possibility of spin $0$ and $1$ mixing. We analyse nature of
the $W$ boson, virtual as well as real, in the last Section.

\section{Generalized solutions of the Dirac equation and mixing of the $S=0$
and $S=1$ spin states}
\label{mixing} 
We have shown that in the non-interacting case full covariant
solutions of the $S=0$ Duffin-Kemmer-Petiau and $S=1$ Hagen-Hurley equations
are generalized (matrix) solutions of the same Dirac equation:%
\begin{equation}
\gamma _{\mu }p^{\mu }\Psi =m\Psi ,  \label{Dirac}
\end{equation}%
see \cite{Okninski2015} for representation of $\gamma ^{\mu }$ matrices.
More exactly, the wave functions $\Psi _{\left( 0\right) }=\left( \psi _{A%
\dot{B}},\ \psi I_{2\times 2}\right) ^{T}$, $\Psi _{\left( 1\right) }=\left(
\zeta _{A\dot{B}},\ -\eta _{C}^{\ D}\right) ^{T}$ $\left( \eta _{CD}=\eta
_{DC}\right) $, are solutions of Dirac equation (\ref{Dirac}) and correspond
to $S=0$,$\ S=1$ cases, respectively \cite{Okninski2015}.

It follows that a function
\begin{equation}
\Psi =c_{1}\Psi _{\left( 1\right) }+c_{0}\Psi _{\left( 0\right) }
\label{mixture}
\end{equation}
fulfills Eq. (\ref{Dirac}) and thus describes a mixture of spin $0$ and spin 
$1$ bosonic states.
Moreover, transformation of spin
states, $S=0\rightleftarrows S=1$, can be constructed explicitly. An
operator $Q$ transforming function $\Psi_{\left( 0\right) }$ into $\Psi
_{\left( 1\right) }$ is: 
\begin{subequations}
\label{QQ}
\begin{gather}
Q\Psi _{\left( 0\right) }=\Psi _{\left( 1\right) }\hspace{58pt}\smallskip
\label{Q1} \\
Q=\left( 
\begin{array}{ll}
I_{2\times 2} & 0_{2\times 2} \\ 
0_{2\times 2} & -\psi ^{-1}\eta _{A}^{~B}%
\end{array}%
\right)  \label{Q2}
\end{gather}
\end{subequations}
where $I_{2\times 2}=\left( 
\begin{array}{cc}
1 & 0 \\ 
0 & 1%
\end{array}%
\right) $, $0_{2\times 2}=\left( 
\begin{array}{cc}
0 & 0 \\ 
0 & 0%
\end{array}%
\right) $, $\eta _{A}^{~B}=\left( 
\begin{array}{cc}
\eta & \eta _{22} \\ 
-\eta _{11} & -\eta%
\end{array}%
\right) $ and $\eta \equiv \eta _{12}=\eta _{21}$. Obviously, we have to
demand also $\zeta _{A\dot{B}}=\psi _{A\dot{B}}$.

The inverse transformation can be also computed. Indeed, we if demand $Q\Psi
_{\left( 1\right) }=Q^{2}\Psi _{\left( 0\right) }$ then we get: 
\begin{equation}
Q^{2}=\left( 
\begin{array}{ll}
I_{2\times 2} & 0_{2\times 2} \\ 
0_{2\times 2} & \psi ^{-2}\left( \eta ^{2}-\eta _{11}\eta _{22}\right)
I_{2\times 2}%
\end{array}%
\right)  \label{Q^2}
\end{equation}%
and thus we also have: 
\begin{equation}
Q\Psi _{\left( 1\right) }=\Psi _{\left( 0\right) }  \label{Q3}
\end{equation}%
provided that $\det \left( \eta _{A}^{~B}\right) =-\eta ^{2}+\eta _{11}\eta
_{22}=-\psi ^{2}$.

\section{$W$ boson in beta decay}
\label{beta}

Interestingly, there are two mechanisms of beta decay, Gamow-Teller (GT) and
Fermi (F). In some cases, known as mixed beta decay, both mechanisms are
present. For example, decay of a neutron is a mixed decay \cite{Krane1988}:%
\begin{equation}
n\longrightarrow p+W^{-}\longrightarrow p+e+\bar{\nu}_{e}  \label{neutron1}
\end{equation}%
More exactly, there are two possible alignments of the electron and neutrino
spins:%
\begin{equation}
n\left( \uparrow \right) \longrightarrow \left\{ 
\begin{array}{l}
p\left( \downarrow \right) +\left[ e\left( \uparrow \right) \bar{\nu}%
_{e}\left( \uparrow \right) \right] \qquad \text{GT transition (}82\text{%
\%)\smallskip } \\ 
p\left( \uparrow \right) +\left[ e\left( \uparrow \right) \bar{\nu}%
_{e}\left( \downarrow \right) \right] \qquad \text{F transition (}18\text{\%)%
}%
\end{array}%
\right.  \label{neutron2}
\end{equation}%
where products of the $W^{-}$ boson decay (see \cite{Tanabashi2018}) are
shown in square brackets and $\left( \uparrow \right) $ denotes spin $\frac{1%
}{2}$.

This is puzzling because the $W$ boson has spin $1$. However, the $W$ boson
is a virtual particle in beta decay and thus spin conservation is open to
discussion. It may be thus assumed that in the case of Fermi decay the $W$
boson is in $S_{z}=0$ state (longitudinal boson) and behaves as a spin $0$
particle. Similar explanation applies to the pion decay \cite{McDonald2016}.

We have proposed alternative explanation, based on the analysis of
Hagen-Hurley equations, suggesting that spin of a virtual $W$ boson belongs
to the $0\oplus 1$ space \cite{Okninski2016}.

\section{$W$ boson in top quark decay}
\label{top}

It can be expected that decay of the top quark may cast new light on nature
of the $W$ boson. The top quark, being the heaviest known fundamental
particle, decays with formation of a real spin $1$ $W$ boson. Moreover, it
decays before hadronization process can occur and thus spin information of
the top quark is not destroyed \cite{Thomson2013,Tsinikos2013}. It can be
shown that in the decay $t\longrightarrow bW\longrightarrow bl\nu _{l}$ the $%
b\nu _{l}$ subsystem is in a $J=0$ state and thus the entire spin of the top
quark is transferred to the lepton \cite{Godbole2019}. Therefore, the
following process should occur exclusively:

\begin{equation}
t\left( \uparrow \right) \longrightarrow b\left( \downarrow \right) +W\left(
\Uparrow \right) \longrightarrow b\left( \downarrow \right) +\left[ l\left(
\uparrow \right) +\nu _{l}\left( \uparrow \right) \right] ,  \label{top1}
\end{equation}%
where the $W$ boson has spin $1$, denoted as $\left( \Uparrow \right) $, see
Section 15.1.1 in \cite{Thomson2013} or Section 2.2 in \cite{Tsinikos2013}.

Suppose now that the $W$ boson is described by the Hagen-Hurley equations.
We have shown that decay of the Hagen-Hurley boson requires that its spin
belongs to $0\oplus 1$ space \cite{Okninski2016}. It follows from results
obtained in Section \ref{mixing} that spin $0$ and spin $1$ states can mix.
Therefore, it is possible that the decaying $W$ boson has not fully
determined spin $S$, i.e. $S\in 0\oplus 1$, and can decay also in the $S=0$
state analogously to the Fermi transition, where $\left(\bigcdot \right) $  below denotes spin $0$:
\begin{equation}
t\left( \uparrow \right) \longrightarrow b\left( \uparrow \right) +W\left(
\bigcdot \right) \longrightarrow b\left( \uparrow \right) +\left[ l\left(
\uparrow \right) +\nu _{l}\left( \downarrow \right) \right].  \label{top2}
\end{equation}
\section{Discussion}
\label{discussion}
Our considerations are based on assumption that the $W$ boson obeys the
(spin $1$) Hagen-Hurley equations. This is possible since these equations do not
conserve parity and can thus describe a weakly interacting particle. 
We have demonstrated that such particle can decay into a lepton
and neutrino but its spin is only partly determined, belonging to the $
0\oplus 1$ space. Furthermore, it follows from Section \ref{mixing} that spin $1$ and spin $0$ 
states can mix, see Eq. (\ref{mixture}).
\medskip

\noindent There are two cases of the $W$ boson decay.
\vspace{-5pt}
\begin{enumerate}
\item When the $W$ boson is virtual (as in the pion decay or in beta decay)
its spin is not conserved. Therefore, there are two mechanisms of mixed beta
decay, cf. Eq. (\ref{neutron2}), moreover the pion decay $\pi ^{+}\longrightarrow
W^{+}\longrightarrow \mu ^{+}+\nu _{\mu }$ is allowed -- this is because
longitudinally polarized $W$ boson behaves as a spin $0$ particle in the
pion decay as well as in the Fermi mechanism of beta decay. We offer
alternative explanation. The virtual $W$ boson is described by the
Hagen-Hurley equations so when it decays its spin is in $0\oplus 1$ space
and hence can decay in spin $0$ state. Ratio of Gamow-Teller to Fermi decay 
is $\left\vert c_{1}\right\vert ^{2}/\left\vert c_{0}\right\vert ^{2}$ with $c_{1}$, $c_{0}$ 
defining the linear combination $\Psi $, see (\ref{mixture}).
\vspace{-5pt}
\item When the $W$ boson is real (as in the top quark decay) the spin is
conserved and, according to the Standard Model, the decay should proceed via
channel (\ref{top1}) only. More exactly, in about $30$\% of the decays the $W
$ boson has negative helicity, in $70$ \% of the events it has longitudinal
helicity, while positive helicity is negligible \cite{Thomson2013}. We
propose alternative possibility. The real $W$ is described by the
Hagen-Hurley equations and due to mechanism of decay (which requires that its
spin belongs to $0\oplus 1$ space \cite{Okninski2016}) and\ mixing of spin $0$
and spin $1$ states, cf. Section \ref{mixing}, decays as a linear combination $\Psi$, cf.  (\ref{mixture}). 
In this case both channels (\ref{top1}), (\ref{top2}) should be observed.
\end{enumerate}

\noindent We assume that decay of the $W$ boson (both
real and virtual) is described by the Hagen-Hurley equations. 
If confirmed, this would
cast new light on mixed beta decay and on the pion decay, as well as on the
top quark decay. Measurement of spins of the products in reaction $%
t\longrightarrow bW\longrightarrow bl\nu _{l}$, especially spin correlation
of $t$ and $b$ quarks,  see Eqs. (\ref{top1}), (\ref{top2}), 
may confirm or rule out this hypothesis.


\begin{thebibliography}{9}
\bibitem{Okninski2016} A. Okni\'{n}ski, Generalized solutions of the Dirac
equation, W bosons, and beta decay, \textit{Adv. High Energy Phys.} 
2016 (2016) Article ID 2689742.

\bibitem{Okninski2015} Andrzej Okni\'{n}ski, Synthesis of relativistic
wave equations: the noninteracting case, \textit{Adv. Math. 
Phys.} 2015 (2015) Article ID 528484.

\bibitem{Krane1988} K.S. Krane, \textit{Introductory Nuclear Physics}, John
Wiley \& Sons, New York, 1988.

\bibitem{Tanabashi2018} M. Tanabashi et al., Review of Particle Physics
(Particle Data Group), \textit{Phys. Rev.} D 98  (2018) 030001.

\bibitem{McDonald2016} K.T.~McDonald, Does Charged-Pion Decay Violate
Conservation of Angular Momentum?, %
\url{http://physics.princeton.edu/~mcdonald/examples/pidecay.pdf}, 2016.

\bibitem{Thomson2013} M. Thomson, \textit{Modern particle physics},
Cambridge University Press, New York, NY, USA, 2013.

\bibitem{Tsinikos2013} I. Tsinikos, Spin correlations at hadron colliders,
MSc Thesis, Universiteit van Amsterdam, 2013.

\bibitem{Godbole2019} R. M. Godbole, M. E. Peskin, S. D. Rindani, R. K.
Singh, Why the angular distribution of the top decay lepton is unchanged by
anomalous $tbW$ couplings, \textit{Phys. Lett.} B 790 (2019) 322-325.
\end{thebibliography}
\end{document}